# Tuning Negative Differential Resistance in a Molecular Film

M. Grobis, A. Wachowiak, R. Yamachika, and M. F. Crommie

Department of Physics, University of California at Berkeley, Berkeley, California 94720-7300 and

Materials Sciences Division, Lawrence Berkeley Laboratory, Berkeley, California 94720-7300

We have observed variable negative differential resistance (NDR) in scanning tunneling spectroscopy measurements of a double layer of $C_{60}$ molecules on a metallic surface. Minimum to maximum current ratios in the NDR region are tuned by changing the tunneling barrier width. The multi-layer geometry is critical, as NDR is not observed when tunneling into a $C_{60}$ monolayer. Using a simple model we show that the observed NDR behavior is explained by voltage-dependent changes in the tunneling barrier height.

Negative differential resistance (NDR) is a crucial property of several important electronic components [1,2]. Originally observed in highly doped tunneling diodes [3], NDR has been seen in a variety of systems and caused by several different mechanisms [4,5,6,7,8]. Here we present a scanning tunneling spectroscopy (STS) study showing the appearance of NDR in the tunneling signature of thin molecular $C_{60}$ films deposited on Au(111). NDR is completely absent for tunneling into a single $C_{60}$ monolayer, but emerges when tunneling into second and higher layers of $C_{60}$. In previous STS studies of molecular systems NDR has been commonly attributed to the convolution of energetically localized tip states with the molecular density of states [7]. The NDR observed in our study is inconsistent with this interpretation, but instead stems from the voltage dependence of the tunneling barrier height [4]. We further find that the relative decrease in current, induced by the NDR, increases with increasing tunneling barrier width, allowing for tunability of the NDR behavior. This behavior is explained by using a simple tunneling model.

Our experiments were conducted using a homebuilt ultrahigh vacuum (UHV) STM with a PtIr tip. The single-crystal Au(111) substrate was cleaned in UHV and dosed with $C_{60}$ using a calibrated Knudsen cell evaporator before being cooled to 7K in the STM stage. dI/dV spectra and images were measured through lock-in detection of the ac tunneling current driven by a 451Hz, 10mV (rms) signal added to the junction bias under open-loop conditions (bias voltage here is defined as the sample potential referenced to the tip). All data were acquired at 7K.

Figure 1(a) shows the topographic structure of a single layer of $C_{60}$ (monolayer), a second layer of $C_{60}$ (bilayer), and a third layer of $C_{60}$ (trilayer). Each layer is well ordered and has a topographic structure consistent with previous measurements performed on similar monolayer and layered $C_{60}$ systems [9,10]. The step height of each $C_{60}$ layer is ~8.0Å. Step edges in the underlying Au(111) substrate lead to 2Å steps that run through the $C_{60}$ layers.

dI/dV spectra performed on the $C_{60}$ monolayer and bilayer are shown in Fig. 1(b). These spectra exhibit several common features: a shoulder in the filled density of states (V<0) and two peaks in the empty density of states (V>0) that arise from tunneling into the $C_{60}$ highest occupied molecular orbital (HOMO), lowest unoccupied molecular orbital LUMO, and LUMO+1, respectively [11]. In the monolayer (bilayer) spectra, the features are centered at -1.4V (-2.0V), 0.75V (1.45V) and 2.05V (2.8V), respectively. There are four important differences between the monolayer and second layer spectra. In comparison to the monolayer spectrum, the bilayer spectrum shows 1) a 1.3V increase in the HOMO-LUMO energy gap [12], 2) a true zero in dI/dV between the HOMO and LUMO edges, 3) 40% narrower molecular resonance widths [13], and 4) the appearance of NDR around for $1.5V < V_{sample} < 2.0V$. In addition, the $C_{60}$ LUMO and LUMO+1 resonances show pronounced splitting in the bilayer spectrum.

The relative magnitude of the current drop due to the bilayer NDR increases with increasing tunnel barrier width. Figure 2(a) shows I-V curves for tunneling into the bilayer at three different barrier widths. The tip height was stabilized at 2.5V with tunneling currents of 5, 50, and 500 pA for these I-V curves, corresponding to junction resistances of 5, 50, and 500GΩ, respectively (each decade increase in tunnel resistance corresponds to approximately a 1Å increase in barrier width). In each I-V trace the current has a local maximum ($I_{max}$) between 1.6V and 1.8V and then decreases, hitting a local minimum ($I_{min}$) between 2.3 and 2.6V before increasing again. The overall fractional change in current ($I_{max}/I_{min}$) due to NDR increases with increasing tunneling barrier width, rising from 1.27 for 5 GΩ to 1.45 for 500 GΩ. The regions of strongest NDR on the molecular film are spatially correlated with the regions of highest $C_{60}$ LUMO state density at the sites of individual bilayer molecules. This can be seen in Fig. 2(b), where red regions in the 1.3V dI/dV map (high LUMO state density) are seen to correlate with dark regions in the 1.9V dI/dV map (high NDR). The solid white circles in these images show the positions of individual $C_{60}$ molecules in the bilayer.

The spatial correlation between LUMO state density and NDR suggests that the observed NDR stems from an inhibition of tunneling to $C_{60}$ bilayer LUMO resonances with increasing voltage. Two mechanisms could produce such behavior: (1) convolution of an energetically localized state on the STM tip with the bilayer density of states [7] and (2) a voltage-dependent increase in the tunneling barrier height [4]. Mechanism 1 can be ruled out since such a mechanism would produce anomalous tip-related spectral features in the $C_{60}$ monolayer spectrum, as well as monolayer NDR. Since neither of these are seen, mechanism 2 remains the most likely candidate for the observed bilayer NDR.

To understand how a voltage-dependent tunnel barrier can lead to NDR in a molecular layer, we simulated our experiment using a basic tunneling formalism for T = 0 K [14]. The model is shown schematically in Fig. 3(a). The tunnel current between the STM tip and $C_{60}$ bilayer can be expressed as

$$I(V,z) \sim \int_0^{|q|_e V} dE\, \rho_{tip}(E_F + E - |q_e|V) \cdot \rho_{sam}(E_F + E) \cdot |T_{tip-sam}(E,V,z_0)|^2 \qquad (1)$$

Here $q_e$ is the electron charge, $|T_{tip-sam}(E,V,z_0)|^2$ is the tunneling matrix element, $z_0$ is the tunnel barrier width (i.e., tip-sample separation), and $\rho_{tip}$ and $\rho_{sample}$ are the tip and sample density of states, respectively. The tunneling matrix element can be estimated using the WKB approximation:

$$|T_{tip-sam}(E,V,z_0)|^2 \sim \exp(-\kappa \int_0^{z_0} dz \sqrt{\phi_0 - E + |q_e|Vz/z_0}) \sim \exp(-\kappa z_0 \sqrt{\phi_0 - E + |q_e|V/2}) \quad (2)$$

where, $\kappa = 2(2m_e/\hbar^2)^{1/2} = 1.02$ Å$^{-1}$eV$^{-1/2}$, and $\Phi_0$ is the work function. We assume the tip density-of-states is constant in the energy window of interest and model the $C_{60}$ density-of-states by four molecular resonances obtained from the observed split $C_{60}$ LUMO and LUMO+1 resonances (splitting is not shown in Fig. 3(a)). The energy width and location of each resonance is estimated from the experimental dI/dV curves. The resonance locations shift slightly with tip-height due to band-bending, an effect that can be estimated from the tip-height dependence of the data. $z_0$ is estimated from the experimental tunneling resistance, R(V), by the relation $z_0 \sim \ln(R(V)/R_0)/\kappa \sqrt{\phi_0 - |q_e|V/2}$ (this can be obtained from Eqs. (1) and (2)). In our simulation we use the $C_{60}$ workfunction (4.7eV) for $\Phi_0$ and assume a "contact resistance" of $R_0 = 100$ kΩ for $z_0 = 0$.

The simulated I-V curves using this model are shown in Fig. 3(b). These plots nicely reproduce the observed NDR magnitude, as well as the observed dependence on tunneling barrier width. The qualitative behavior of the calculated I-V curves is very robust and is not sensitive to the exact values used for orbital energies, band-bending, and tip-sample separation. The NDR behavior can be qualitatively explained by considering the tunnel current into a single energetically localized resonance at energy $E_0 = V_0|q_e|$. For $V > V_0$, the ratio $I(V) / I(V_0)$ can be derived from Eqs. (1) and (2):

$$I(V) / I(V_0) = \exp(-\kappa z_0 (\sqrt{\phi_0 - E_0/2 + |q_e|V/2} - \sqrt{\phi - E_0/2})) \quad (3)$$

This ratio is always less than one and NDR is thus always observed for $V > V_0$ in this case. Due to the multiplicative factor of $z_0$ in the exponent, $I(V)/I(V_0)$ decreases more rapidly with V for larger $z_0$, enhancing NDR for greater tip-sample separation. Thus, by varying $z_0$, the degree of NDR can be precisely tuned. NDR is completely eliminated, however, if a sufficiently large constant density of states is included in addition to the narrow resonance. Conduction through such a metallic continuum overpowers the current drop due to voltage dependent barrier and prevents NDR from emerging. This explains why NDR is not observed in the monolayer spectra. The $C_{60}$ monolayer has an increased interaction with the Au(111) substrate, which adds a constant metallic density of states background to the monolayer electronic structure. Increased substrate interaction also broadens the $C_{60}$ molecular resonance widths [11,15,16]. The combination of these factors suppresses NDR from emerging in $C_{60}$ monolayer tunneling.

In conclusion, we have observed the emergence of tunable NDR in tunneling to a $C_{60}$ bilayer. The observation of NDR in this experiment was dependent on three key factors: (1) a voltage-dependent tunneling barrier, (2) absence of a broad metallic density of states, and (3) narrow tunneling resonances. NDR arising from this mechanism increases with increased tip-sample separation, allowing for the values of $I_{max}/I_{min}$ to be precisely tuned.

This work was supported in part by NSF Grant No. EIA-0205641 and by the Director, Office of Energy Research, Office of Basic Energy Science, Division of Material Sciences and Engineering, U.S. Department of Energy under contract No. DE-AC03-76SF0098.

**Figure Captions**

FIG.1 (a) 300Åx300Å topograph (V = 2.5V, I = 20pA) of the $C_{60}$ monolayer, second layer (bilayer), and third layer (trilayer) regions on the Au(111) surface. (b) dI/dV spectra measured on the $C_{60}$ monolayer (black curve) and $C_{60}$ bilayer (blue curve). NDR regions in bilayer dI/dV can be seen for 1.6V < V < 2.5V. The curves are scaled by normalizing to the LUMO+1 maximum. Tip stabilization: 2.5V, 200pA (monolayer) and 3.5V, 100pA (bilayer).

FIG. 2 (a) I-V curves acquired at three different tunnel junction resistances on the $C_{60}$ bilayer. Each curve is normalized to the respective local current maximum ($I_{max}$) located at 1.5V < $V_{max}$ < 2.0V. Tip stabilization: V = 2.5V; I = 5pA, 50pA, 500 pA for the black, red, and blue curves, respectively. (b) Topograph (V = 3.1V, I =10 pA) and energetically resolved dI/dV maps for $C_{60}$ bilayer taken at voltages just below and above $V_{max}$. The tip height was stabilized at 3.1V, 100pA for each dI/dV map. Superimposed white circles show the locations of individual $C_{60}$ bilayer molecules. The color scale applies only to the dI/dV maps.

FIG. 3 (a) Schematic of tunnel junction model used in simulating $C_{60}$ bilayer I-V. The STM tip is separated by a distance $z_0$ from the $C_{60}$ bilayer and the tip density of states is assumed to be constant. Splitting in the bilayer LUMO (L) and LUMO+1 (L+1) resonances is not shown in the sketch. (b) Simulated I-V curves corresponding to the three experimental I-V curves shown in Fig. 2. NDR occurs for V > $V_{max}$ due to increased effective barrier for tunneling into LUMO resonances. Increase in NDR with tip-sample separation is well reproduced in this model.

Figure 1

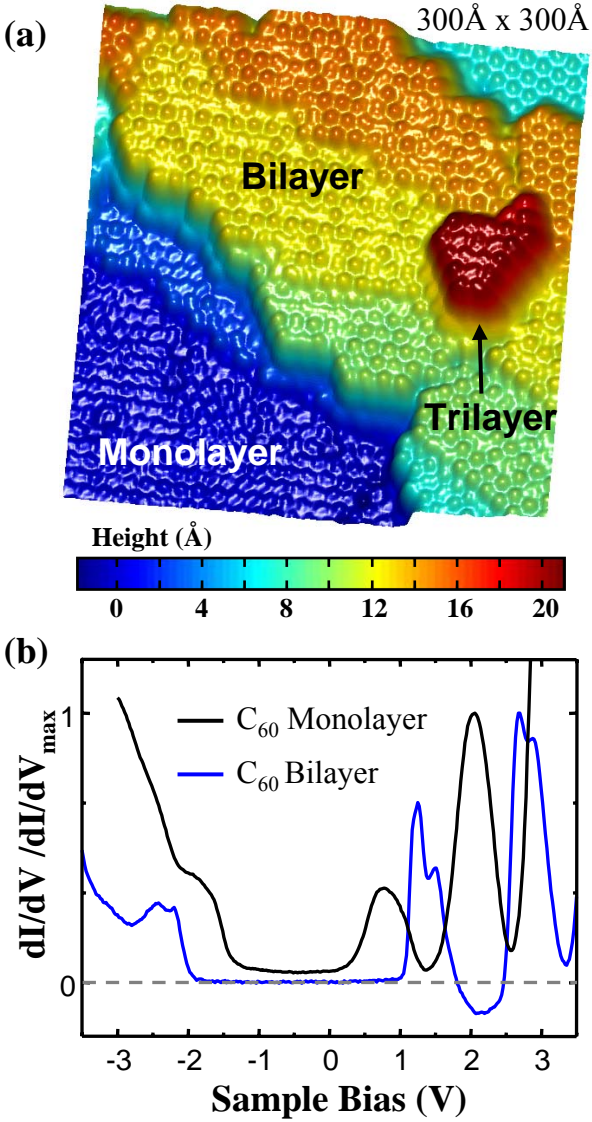

Figure 2

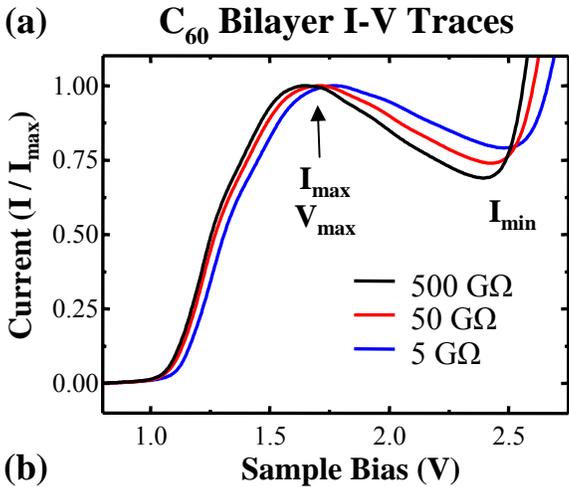
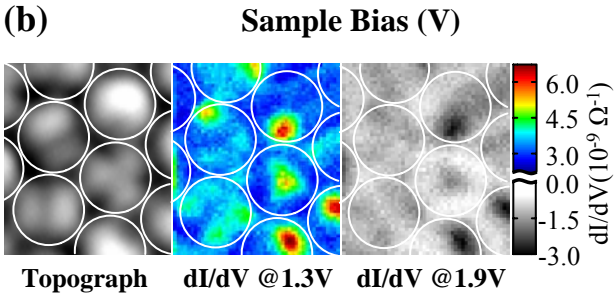

Figure 3

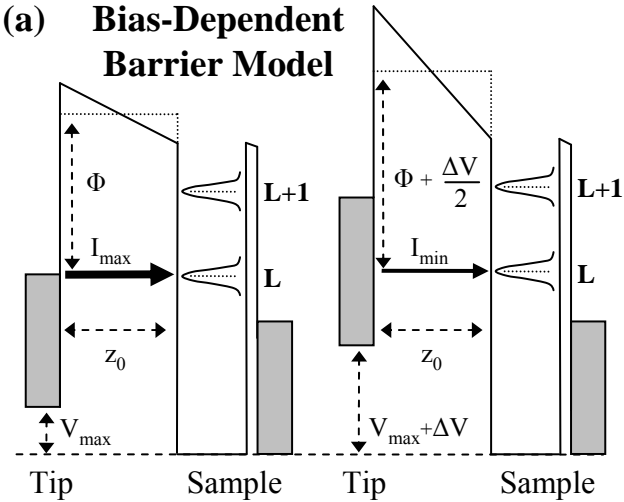

(a) **Bias-Dependent Barrier Model**

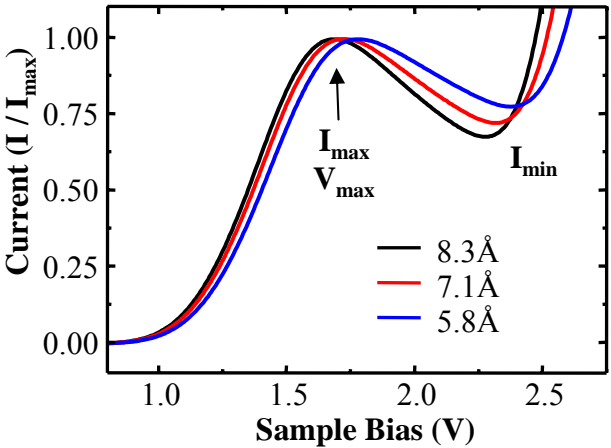

(b) **Simulated I-V for $C_{60}$ Bilayer**